\documentclass{article}
\usepackage{spconf,amsmath,graphicx}
\usepackage{epstopdf}
\usepackage{multirow}
\usepackage{multirow, makecell}
\usepackage{amsmath}
\usepackage{amssymb}

\makeatletter
\def\set@curr@file#1{%
  \begingroup
    \escapechar\m@ne
    \xdef\@curr@file{\expandafter\string\csname #1\endcsname}%
  \endgroup
}
\def\quote@name#1{"\quote@@name#1\@gobble""}
\def\quote@@name#1"{#1\quote@@name}
\def\unquote@name#1{\quote@@name#1\@gobble"}
\makeatother

\title{Decoding Movement Imagination and Execution from EEG Signals using BCI-Transfer Learning Method based on Relation Network}

\name{Do-Yeun Lee$^1$ \qquad Ji-Hoon Jeong$^1$ \qquad Kyung-Hwan Shim$^1$ \qquad Seong-Whan Lee$^{1,2}$\thanks{This work was partly supported by Institute of Information \& Communications Technology Planning \& Evaluation (IITP) grant funded by the Korea government (No. 2017-0-00432, Development of Non-Invasive Integrated BCI SW Platform to Control Home Appliances and External Devices by User’s Thought via AR/VR Interface) and partly funded by Institute of Information \& Communications Technology Planning \& Evaluation (IITP) grant funded by the Korea government (No. 2017-0-00451, Development of BCI based Brain and Cognitive Computing Technology for Recognizing User’s Intentions using Deep Learning).}}
\address{$^1$Department of Brain and Cognitive Engineering, Korea University\\$^2$Department of Artificial Intelligence, Korea University}

\begin{document}
%
\maketitle
\begin{abstract}
A brain-computer interface (BCI) is used not only to control external devices for healthy people but also to rehabilitate motor functions for motor-disabled patients. Decoding movement intention is one of the most significant aspects for performing arm movement tasks using brain signals. Decoding movement execution (ME) from electroencephalogram (EEG) signals have shown high performance in previous works, however movement imagination (MI) paradigm-based intention decoding has so far failed to achieve sufficient accuracy. In this study, we focused on a robust MI decoding method with transfer learning for the ME and MI paradigm. We acquired EEG data related to arm reaching for 3D directions. We proposed a BCI-transfer learning method based on a Relation network (BTRN) architecture. Decoding performances showed the highest performance compared to conventional works. We confirmed the possibility of the BTRN architecture to contribute to continuous decoding of MI using ME datasets.
\end{abstract}
 

\section{Introduction}
\label{sec:intro}
Brain-computer interfaces (BCI) can be used as communication systems that translate the user's intentions into commands to control external devices. BCI technology has adapted assisted devices \cite{jeong2019trajectory}, to rehabilitate a functional recovery for spinal cord injury (SCI) and stroke patients, while also providing support in multiple tasks encountered in daily life \cite{penaloza2018bmi, jeong2020decoding}. Electroencephalogram (EEG) is one of the non-invasive methods to obtain brain signals, and is commonly used to construct an effective BCI system due to the high temporal resolution and low cost. Movement imagination (MI) paradigm has endogenous characteristics because it does not need an external stimulus to perform the given task \cite{townsend2004continuous}. Therefore, it could be adopted for a more intuitive brain-to-robot interaction compared to other BCI paradigms \cite{ang2015randomized}. However, recent MI studies have not achieved sufficient decoding performance. The performance of the MI task often increases participant's fatigue, and it is difficult to determine which movement is exactly imagined. In contrast, a movement execution (ME) task is easier to perform and actually uses the same limbs, making the brain signal more consistent. Additionally, acquired signals during the ME task showed a stronger amplitude characteristics compared to MI task \cite{miller2010cortical}.

Therefore, in this study, we adopted the principle of transfer learning for a robust MI decoding using the ME task dataset as a supporting tool. Recent studies proposed advanced decoding methods based on deep learning and transfer learning for successful decoding of user-intuitive intention. Samek et al. \cite{samek2013transferring} proposed a method for transferring nonstationary information between subjects, which effectively narrows the gap between training data and test data. Azab et al. \cite{azab2019weighted} applied transfer learning in the classification domain, which maintained classification performance even when there were few subject-specific trials available for training. Fahimi et al. \cite{fahimi2018inter} performed inter-subject transfer learning methods on a classification domain for an end-to-end deep CNN help to decode the attentional information from time-series. Tan et al. \cite{tan2019attention} presented an attention-based transfer learning method, applying it to the EEG classification domain. They also proposed EEG optical flow and deep transfer learning that is appropriate for transferring knowledge through joint training \cite{tan2018deep}. Giles et al. \cite{giles2019subject} proposed a subject-to-subject transfer learning framework. They suggested a new measurement named the Jensen Shannon Ratio (JSR) that is used to compare calibration trials with existing data sets for transfer learning.  

In this paper, we focused on the decoding of the various types of upper extremity movements from EEG signals. Especially, arm reaching tasks in multiple directions, which are important since it is how any kind of upper extremity movement starts. In order to decode various upper limb movements, we acquired EEG data using six arm reaching tasks (left, right, forward, backward, up and down) for real and imaginary movements. We proposed BCI-transfer learning based on the Relation network (BTRN) using the MI and ME paradigm for robust decoding of various upper limb movements. Using the ME paradigm can reduce participant fatigue and have more consistent and robust signal characteristics than using the MI paradigm. Therefore, using the real movement can be applied as an advantage to replace a part of the MI paradigm. To the best of our knowledge, this is the first attempt to applying transfer learning for EEG data of high-complexity tasks. We demonstrated the possibility of the BTRN architecture to contribute in decoding an MI paradigm using ME datasets.

\section{Experiments}
\label{sec:typestyle}

\subsection{Participants}
\label{ssec:subhead}

Nine healthy and right-handed participants (4 females and 5 males, 22-27 years) were recruited for the experiment. None of them had prior knowledge of BCI experiments. Before the start of the experiment, we explained the entire experimental protocols to all participants. The protocols and environments were reviewed and approved by the Institutional Review Board at Korea University [1040548-KU-IRB-17-172-A-2].

\subsection{Experimental protocol}
\label{ssec:subhead}

Participants sat comfortably in chairs, with visual displays in front of them. The experiment comprised of two sessions: an executive session and an imaging session. Participants were asked to perform the arm reaching tasks for six different directions (left, right, forward, backward, up and down), respectively (Fig. 1). The order of tasks were cued randomly. In the motor execution session, participants performed real movements of arm reaching tasks according to the visual cue. For the imagination session, they imagine the shown task without any movement. The experiment paradigm consists of three seconds of rest, three seconds of visual instruction and four seconds of ME or MI tasks. We recorded 300 trials for each session and 50 trials per directions. We recorded EEG signals with 60 electrodes, reference and ground channels were placed on FCz channel and Fpz according to the international 10/20 system respectively. EEG signals were recorded with the ActiCap system, BrainAmp amplifiers (Brain Product GmbH, Germany) and a MatLab 2018a software.  

\begin{figure}[t!]
\centering
\centerline{\includegraphics[width=\columnwidth]{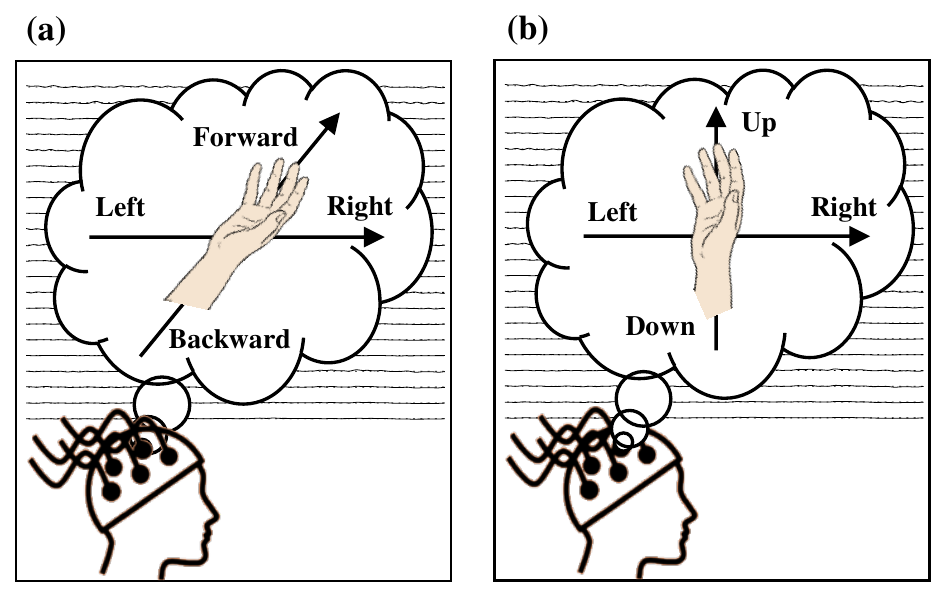}}
\caption{Experimental protocols for arm reaching tasks. (a) horizontal direction and (b) vertical direction}
\label{fig:res}
\end{figure}   
\begin{figure*}[t!]
\centering
\centerline{\includegraphics[scale = 0.85]{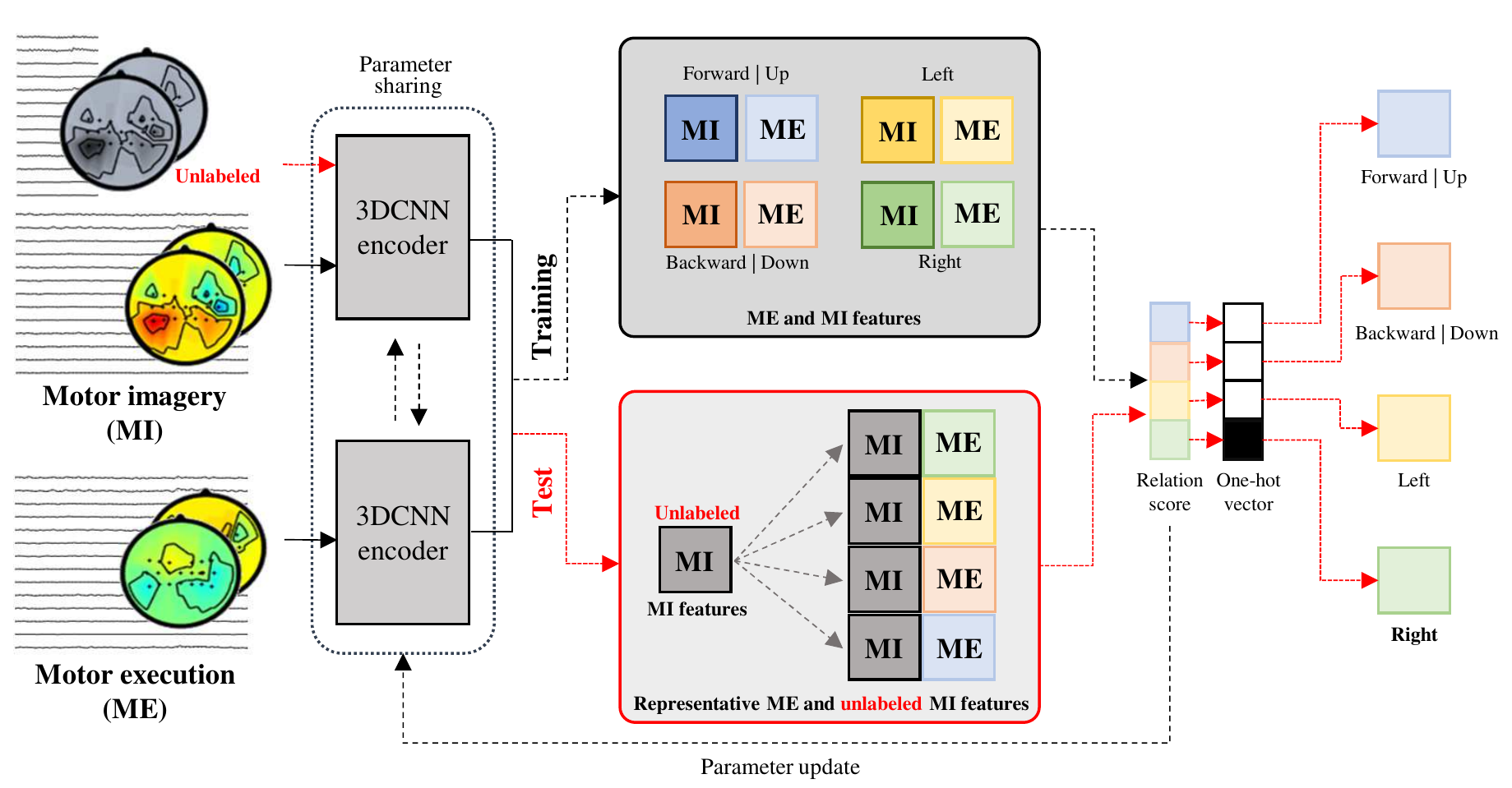}}
\caption{Overall flowchart of the proposed BTRN architecture using MI and ME dataset}
\label{fig:res}
\end{figure*}

\section{Methodology}
\label{sec:pagestyle}

\subsection{Data preprocessing}
\label{ssec:subhead}
The acquired EEG signals were down-sampled from 1000 Hz to 250 Hz. A bandpass filtering of 4-40 Hz was applied using a zero phase, fifth-order Butterworth filter. Also common average referencing (CAR) based spatial filtering to enhance signal quality was used. We analyzed three seconds of data in each ME and MI session. We segmented the entire filtered signals into 80\% training data and 20\% test data, respectively. We selected and analyzed only 25 channels (F3, F1, Fz, F2, F4, FC3, FC1, FCz, FC4, C3, C1, Cz, C2, C4, CP3, CP1, CPz, CP2, CP4, P3, P1, Pz, P2 and P4). Channels located in the prefrontal sides were not used to avoid artifacts caused by eye movement \cite{chang2016detection}. Channels in temporal and occipital areas were also not used because they are known to be related to sound and visual inspection \cite{romei2012sounds}.

\subsection{BCI-Transfer learning method based on Relation Network (BTRN)}
\label{ssec:subhead}

We propose a BTRN architecture that classifies various upper limb movements using both ME and MI datasets (Fig. 2). We utilize a Relation network with a Siamese network in order to apply transfer learning to the MI classification.

\subsubsection{Relation network}
\label{ssec:subsubhead}
Relation networks have been used to classify new classes by calculating the relationship score between the query image and the new class \cite{sung2018learning}. These approaches do not update the network in order to properly classify even small datasets. The fundamental problem of EEG signals in the BCI paradigm is that the data aquisition process is difficult and complex. In general, MI-based BCI collects MI data as well as ME data, however the later one is not used during classification. Therefore, if we can use ME data properly, the traditional BCI problem will be alleviated. Firstly, ME and MI dataset are fed into the Siamese network to obtain the same size of features over the same encoders. The obtained features coming from the same encoder are then concatenated, thereby doubling the features. Through this, we comprise the features of the MI and ME dataset. The Relation network receives the concatenated features as input and conducts convolutions to calculate the similarities between representative ME features and the unlabeled MI features. Using these similarities, the Relation network generates a relation scores. In the last layer, the softmax function scales the relation scores from 0 to 1 for classification. Additionally, we applied average pooling that smoothly interpolates between common pooling operators. Details on the Siamese network are described in in section 3.2.2.

\subsubsection{Siamese network}
\label{ssec:subsubhead}
We used transfer learning with the Siamese network for obtaining the features of the ME and MI dataset. Because this network is trained with two identical encoders, the features of the ME and MI dataset can be updated together. Basically, the Siamese network is intended to learn better features for classification \cite{pan2010survey, koch2015siamese}. In this study, two identical 3D convolution neural network (CNN) architectures are constructed. This architecture considers temporal, spatial and spectral features of brain signals. We rearrange input data from two dimensions (time $\times$ channel) to three dimensions (channel $\times$ channel $\times$ time), in order to have a relation between spatial characteristics and channel position \cite{shim2019assistive}. The 3DCNN encoders are comprised of two main layers, a Conv3D and average pooling layer, to reduce the data significantly. We use both ME data and MI data as inputs for the encoders by concatenating each output feature to construct the data. Since a 3DCNN is used for feature extraction, the output of the network contains features from a combination of ME data and MI data. The output features are sent into the relation network to calculate the relation scores based on similarities. The encoders share the parameters and weights, therefore the encoders extract features that are more suitable for finding similarities as the training progresses.  
\section{Results and discussion}
\label{sec:majhead}

Table 1 indicates the classification accuracies for the arm reaching tasks for all participants using the proposed BTRN. In the horizontal plane, the grand-averaged classification results over all subjects achieved an accuracy of 0.46 ($\pm$0.02) using the combined data (ME and MI data) and 0.45 ($\pm$0.05) using just MI data. Also, in the vertical plane, the proposed model achieved an accuracy on both datasetes of 0.46 ($\pm$0.01) and 0.45 ($\pm$0.04) respectively. The performance of classification in each data was above chance level (0.25). As a result, we showed that using the combined data can provide similar classification performance as using MI data.

Classification comparison performances using common spatial pattern (CSP) and linear discriminant analysis (LDA) \cite{wu2013common}, DeepConv \cite{schirrmeister2017deep}, EEGNet \cite{lawhern2018eegnet}, and proposed BTRN architecture are presented in Table 2. Performance indicates the grand-averaged classification accuracy over all participants. In the horizontal plane, the proposed BTRN architecture has the highest classification accuracies (combined: 0.46, MI: 0.45) while CSP$+$LDA model showed the lowest (combined: 0.31, MI: 0.30). Similarly, in the vertical plane, the highest class accuracies were 0.46 and 0.45 in combined and MI respectively using the BTRN architecture, while the lowest result was 0.29 and 0.33 in each of the datasets using CSP+LDA. This result proves that the BTRN is more robust than other conventional models when performing multiclass-classification.

Fig. 3 shows the confusion matrices of each class for a representative participant sub3 and sub1 in the combined dataset. In the horizontal plane task, classification accuracy was 0.50 but the true positive value for `right' class was 0.42. However, in vertical plane, `left' class had the lowest value (0.42). We also observed that the matrix in vertical plane is mainly confused between `left' and 'right' classes. The highest results in horizontal and vertical planes were in `backward' and `down' class, respectively. The results show that there is a performance difference for each class, but it is confirmed that there is no large variation between sessions. Thus, this result demonstrates that it is possible to decode various reaching movements from the same limb in the vertical and horizontal planes.

\begin{table}[t!]
\small
\caption{Classification accuracies of arm reaching task according to proposed BTRN architecture}
\renewcommand{\arraystretch}{1.3}
\resizebox{\columnwidth}{!}{%
\begin{tabular}{ccccc}\hline
\multirow{2}{*}{Subjects} \multirow{2}{*}{} & \multicolumn{2}{c}{4-class (horizontal plane)} & \multicolumn{2}{c}{4-class (vertical plane)} \\ & ME+MI & MI & ME+MI & MI \\ \hline
sub1 & 0.45 & 0.42 & 0.47 & 0.45 \\
sub2 & 0.42 & 0.45 & 0.45 & 0.40 \\
sub3 & 0.50 & 0.57 & 0.44 & 0.42 \\
sub4 & 0.47 & 0.40 & 0.45 & 0.52 \\
sub5 & 0.48 & 0.50 & 0.50 & 0.47 \\
sub6 & 0.45 & 0.45 & 0.45 & 0.42 \\
sub7 & 0.45 & 0.37 & 0.45 & 0.50 \\
sub8 & 0.45 & 0.47 & 0.47 & 0.42 \\
sub9 & 0.47 & 0.45 & 0.45 & 0.45 \\
\textbf{Average} & \textbf{0.46} & \textbf{0.45} & \textbf{0.46} & \textbf{0.45} \\
Std. & 0.02 & 0.05 & 0.01 & 0.04 \\\hline
\end{tabular}%
}
\end{table}

\begin{table}[t!]
\caption{Comparison performances using the BTRN and the conventional methods}
\renewcommand{\arraystretch}{1.5}
\resizebox{\columnwidth}{!}{%
\begin{tabular}{ccccc}\hline
\multirow{2}{*}{Methods} & \multicolumn{2}{c}{4-class (horizontal plane)} & \multicolumn{2}{c}{4-class (vertical plane)} \\ 
 & ME+MI & MI & ME+MI & MI \\ \hline
CSP$+$LDA \cite{wu2013common} & 0.31 ($\pm$0.04) & 0.30 ($\pm$0.04) & 0.29 ($\pm$0.03) & 0.33 ($\pm$0.04) \\
DeepConv \cite{schirrmeister2017deep} & 0.31 ($\pm$0.03) & 0.36 ($\pm$0.03) & 0.34 ($\pm$0.07) & 0.34 ($\pm$0.05) \\
EEGNet \cite{lawhern2018eegnet} & 0.33 ($\pm$0.05) & 0.35 ($\pm$0.04) & 0.35 ($\pm$0.06) & 0.34 ($\pm$0.04) \\ 
\textbf{BTRN} & \textbf{0.46 ($\pm$0.02)} & \textbf{0.45 ($\pm$0.05)} & \textbf{0.46 ($\pm$0.01)} & \textbf{0.45 ($\pm$0.04)} \\ \hline
\end{tabular}%
}
\end{table}
\begin{figure}[t!]
\centering
\centerline{\includegraphics[scale = 0.85] {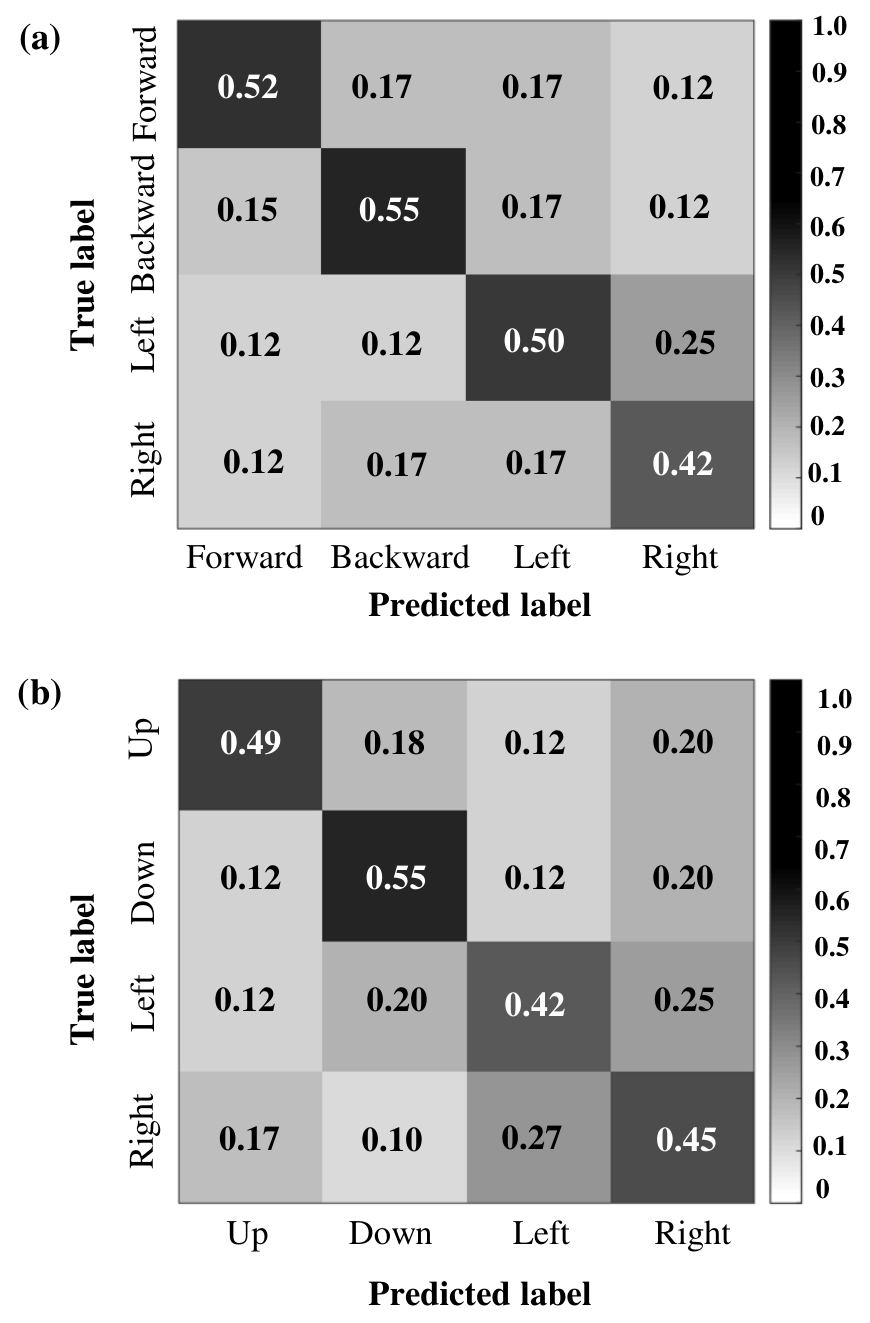}}
\caption{Confusion matrices of each class for the representative subjects, respectively. (a) horizontal plane for sub3 and (b) vertical plane for sub1}
\label{fig:res}
\end{figure}

\section{conclusion and future work}
\label{sec:print}

In this paper, we proposed BTRN architecture and obtained higher multiclass-classification accuracies compared to conventional models. We proved that the feasibility of robust decoding six different arm movements using ME and MI data for transfer learning. Using the ME paradigm can be replaced by some of the MI paradigm and is more consistent in subject and EEG signals. Thus, we demonstrated that BTRN architecture is relatively effective for decoding multiple movement intention from EEG signals. Future work will be to develop robust decoding methods based on deep learning to overcome problems with small EEG dataset. 

\section{Acknowledgement}
The authors thanks to B.-H. Kwon, B.-H. Lee and J.-H. Cho for their help with the dataset construction and S. K. Prabhakar, P. Bertens, and J. Kalafatovich for their useful discussions.

\bibliographystyle{IEEEbib}
\bibliography{strings,refs}

\vfill\pagebreak

\end{document}